# Selective Match-Line Energizer Content Addressable Memory (SMLE-CAM)

Mohammed Zackriya. V and Harish M Kittur, *Member, IEEE*
School of Electronics Engineering
VIT University
Vellore – 632 014, India
mdzackriya@vit.ac.in, kittur@vit.ac.in

*Abstract*— A Content Addressable Memory (CAM) is a memory primarily designed for high speed search operation. Parallel search scheme forms the basis of CAM, thus power reduction is the challenge associated with a large amount of parallel active circuits. We are presenting a novel algorithm and architecture described as Selective Match-Line Energizer Content Addressable Memory (SMLE-CAM) which energizes only those MLs (Match-Line) whose first three bits are conditionally matched with corresponding first three search bit using special architecture which comprises of novel XNOR-CAM cell and novel XOR-CAM cell. The rest of the CAM chain is followed by NOR-CAM cell. The 256 X 144 bit SMLE-CAM is implemented in TSMC 90 nm technology and its robustness across PVT variation is verified. The post-layout simulation result shows, it has energy metric of 0.115 fJ/bit/search with search time 361.6 ps, the best reported so far. The maximum operating frequency is 1GHz.

## I. INTRODUCTION

Content Addressable Memory (CAM) retrieves address of the matched content location in memory against the search data. CAM is capable of high speed search by implementing parallel bit by bit search within one clock cycle [1]. CAM is extensively used in applications like IP routing [2], data compression & image processing viz., Gray coding [3], XML parsing [4], where software search algorithms fail to attain high speed. CAM is power hungry due to parallel data processing; therefore to reduce power and increase speed, many works with different match-line strategies [5-15], have been proposed.

The traditional CAM is designed using either NOR-logic match-line [1] which introduces low search delay with the cost of high power consumption due to high switching activities or NAND-logic match-line [1] which introduces low power consumption with the cost of high search delay because of long pull down path. Thus when the whole CAM is built using NOR-logic, due to drain capacitance it consumes more power for match-line switching since most of the CAM words are mismatched in practical applications. A hybrid CAM is proposed [5], where the high speed search performance of the NOR-logic match-line and the low power performance of the NAND-logic match-line are combined. The hybrid CAM costs area of additional transistors (nine) in control circuitry to overcome race conditions for every wordline in the CAM. [15] proposes an architecture which uses only NOR-logic CAM cells with two stages: (1) Pre-search and (2) Main-search. Similar to [5], the second stage will discharge the ML only if the first stage matches using Early Predict Late Correct (EPLC) architecture.

It is well known that the power consumption $P_{d,\text{ charge/discharge}} = \alpha f C_L V_{DD} V_S$, where $V_s$ is voltage swing of the node. Therefore, the overall power consumption can be reduced by reducing the swing voltage of either match-line or searchline. Swing voltage at match-line is reduced roughly by a factor $2V_{th}/V_{DD}$ [6]. LSSL [7], swing voltage at searchline is $V_L = V_{REF} - \Delta V$ and $V_H = V_{REF} + \Delta V$. Low swing voltage not only reduces power consumption, it also reduces search delay because of lower charging and discharging time. But it requires very sensitive current supplying circuit which increases designing complexity and additional area for every wordline.

Another type of common CAM is precomputation-based CAM (PB-CAM) [8] & [9], where a precomputation stage is introduced before CAM memory. The precomputation stage consists of a parameter extractor and parameter memory. Here a particular parameter (say number of 1's in a word) of the stored words (in CAM) is stored in parameter memory and the incoming search data is passed on to parameter extractor where the parameter of the search word will be extracted and it will be compared to the stored parameter in the parameter memory. It is to be noted that only when the parameters are matched, the data will be sent to the CAM. If the parameters are mismatched then the data will not be sent to CAM for search, thus saving power. The parameter extractor is implemented in different ways using full-Adder [8] and chain of XOR logic [9]. Even though complex algorithms are used to reduce search in second stage but due to precomputation process for every search word in the first stage, the overall search time increases.

To reduce the overall power consumption pipelined hierarchy search scheme is introduced [10], the match-line is broken into several segments and later the search operation is done from the first segment to the last. Suppose mismatch occurs in a segment, then the search operation will be denied for the successive segments. To attain high performance by changing the NAND-logic match-line, an AND type match-line scheme [11] is constructed using PF-CDPD logic. A self disabled sensing technique with differential match-line is used

[12] to choke the charge current to the match-line with a feedback loop, once the matching operation is started. The authors have combined SL (Search Line) hierarchy and a new butterfly match-line scheme to attain energy efficiency [13]. Along with the power-gated ML sensing, in order to give robustness from noise, a parity bit is introduced which reduces the sensing delay [14] but with power overhead.

In this paper we are proposing a novel match-line scheme and CAM architecture with novel CAM cells and the contents are organised as follows: The abstract view of proposed architecture is presented in Section II. Section III describes the design of novel XNOR, novel XOR and NOR CAM cell. Description of Match-Line Energizer (MLE) circuit is given in Section IV. Section V describes the algorithm and architecture of SMLE-CAM. The implementation results are shown in section VI. The performance of SMLE-CAM is compared with other CAM in section VII. Section VIII concludes with the performance characteristics of SMLE-CAM.

## II. BASIC CONCEPT OF PROPOSED CAM ARCHITECTURE

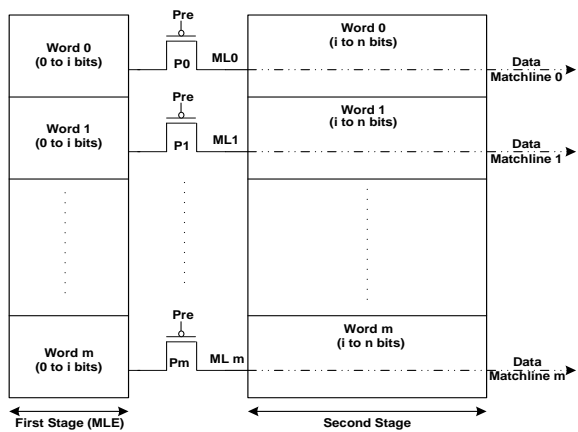

Fig. 1. Conceptual view of proposed CAM Architecture

To the best of our knowledge, the CAM architecture presented in the literature survey have the source of the precharge device (that charges the Matchline) directly connected to the voltage source ($V_{DD}$). In our proposed architecture, the word length (n) is divided into two segments i.e., 0 to i bits and i to n bits as shown in Fig.1. The division ratio is made so as to achieve the best overall energy metric. The first stage is constructed using MLE circuit (as shown in Fig.6), which forms the source for the precharge devices (P0, P1,…, Pm). Thus every ML of a particular wordline is precharged conditionally only if the bits of the corresponding wordline match in the first stage. Since the searching process is performed parallel in both the stages and the ML is precharged conditionally, the performance of the CAM is improved in terms of both search time and power consumption.

## III. DESIGN OF THE CAM CELLS

A CAM is built using number of CAM cells. CAM cells are used to both store and compare the data. Every CAM cell is connected to the ML. Depending upon whether the stored bit in the CAM cell and the search bit matches or mismatches, the CAM cell will either keep ML high or pull down to low respectively. The primary architecture of our proposed CAM differs from the other CAM architecture is that; the CAM has a first stage, the MLE circuit. The CAM architecture consists of novel CAM cells which are, to the best of our knowledge, presented for the first time;

### A. Novel XNOR CAM Cell for first stage

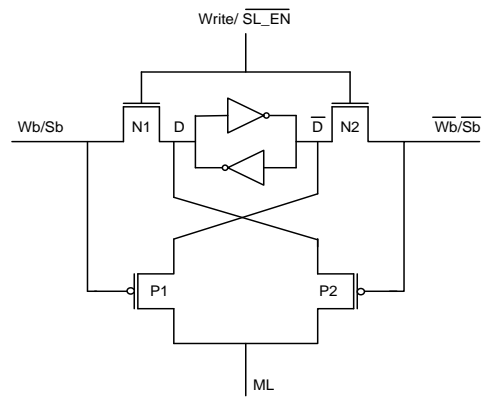

Fig. 2. Novel XNOR CAM circuit

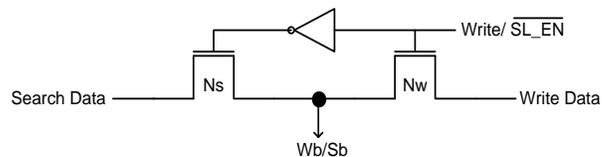

Fig. 3. Write/search driver circuit

The first bit of search word is compared using XNOR CAM cell [Fig. 2]. Match of first bit will form the power source for the entire ML.

*Case1: (Write/$\overline{SL\_EN}$ = 1)*
The driver circuit shown in Fig. 3 enables write operation by turning on Nw, N1 and N2, so that the data can be written into the SRAM latch.

The advantage here is that the same path can be used for both writing and searching data.

*Case2: (Write/$\overline{SL\_EN}$ = 0)*
The driver circuit enables search operation by turning on Ns and it will turn off Nw, N1 and N2 to preserve the data in the latch. When search bit (Sb) and D is same(1), P2 will be on thus enabling power to ML. When Sb and D is same (0), P1 will be on thus enabling power to ML. When Sb and D are different, ML will be held low, thus ML will not be precharged to high.

### B. Novel XOR CAM Cell for first stage

The XNOR CAM cell charges ML (Fig.2) in case of match and no charge yields through ML in case of mismatch. But the second and third bit (which is fed into MLE circuit) has to charge ML in case of mismatch and in case of match no charge should yield though ML. Thus, the second and third bits of search word are compared using XOR CAM cell [Fig. 4].

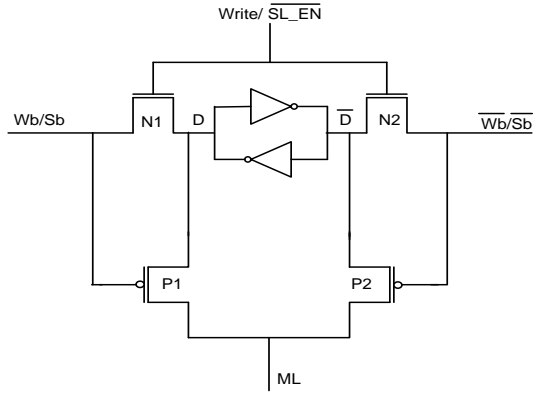

Fig. 4. Novel XOR CAM circuit

Same driver circuit as in Fig.3 is used here for alternatively writing and searching data. Write operation is same as that of the XNOR CAM cell.

During search operation, when Sb and D are same, the ML will be held low. When Sb=0(1) and D=1(0) are different, P1 (P2) will be on and thus ML will be pulled to high.

XNOR and XOR CAM cell form the first stage of the CAM cell.

*C. NOR CAM Cell for second stage*

The second stage (Fig.1) of the search bits are matched using NOR CAM cell as shown in Fig. 5. Write operation takes place similar to SRAM [16] using pass-transistors.

The ML is precharged to high only when first three bits are matched in the first stage (Fig. 1). This precharged ML is connected with the chain of NOR CAM cell. If any bit of the word is mismatched, then the ML will be pulled to low through N1 and N2 (or N3 and N4). Since the pull down transistor pair is connected in NOR fashion, the search performance is very high.

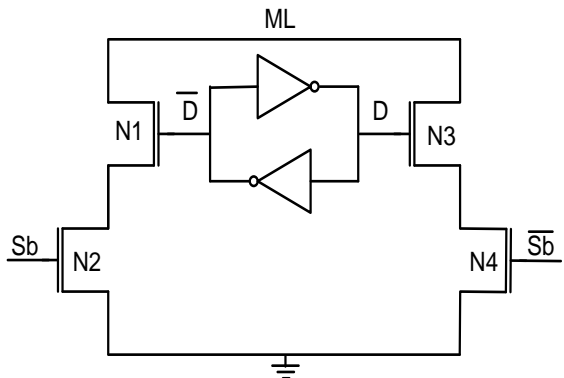

Fig. 5. NOR CAM circuit

IV. MATCH-LINE ENERGIZER

The proposed Match-Line Energizer (MLE) circuit, which forms the primary circuit for our architecture is depicted in Fig. 6.

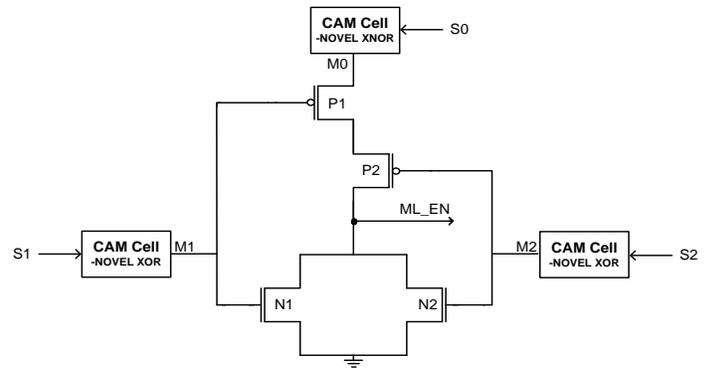

Fig. 6. Match-Line Energizer circuit

The arrangement is similar to NOR implementation. Traditionally power to the ML is supplied by the power source directly [5-15], which enables continuous charge flow to ML unconditionally with respect to PRE signal. Unlike other CAM, in SMLE-CAM, charge will flow into ML only if the first three bits are conditionally matched. Thus it reduces search time and unnecessary charge flow to the remaining $ML_S$ ($M3$-$M_{n-1}$) for 87.5% (7/8*100) of the all possible inputs to the ML ($ML_0$-$ML_{N-1}$) through P1 and P2, when the stored data is distributed uniformly. Here the source for the entire circuit is M0, which is the ML of the XNOR CAM cell. Since the precharge of ML is done through 5 staggered transistors (2 in XNOR CAM cell, 2 in MLE circuit and 1 in precharge device), the devices of bigger size are used so that the CAM differentiates the match and mismatch at 1GHz.

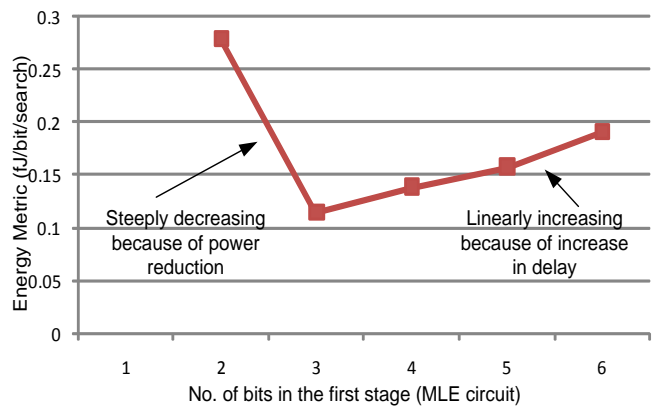

Fig. 7. Energy metric for various bit length of MLE circuit

The MLE circuit is best designed only for three bits of length. In case, if the MLE is designed for more than 3 bits; it will result in reduced power (as only less number of MLs will be energized), but the overhead in search delay increases, which ultimately degrades the energy metric (in Fig.7, the energy metric is increasing linearly from 3 to 6 bits). The primary reason for this is, the number of PMOS in series will increase with increase in bit length of MLE and will create charge flow problem from M0 to ML_EN. If the length of bit is reduced from three (i.e. two bits), then again the CAM consumes more power comparatively (with 3 bit MLE) because it reduces the unnecessary charge flow to the MLs only for 75% (3/4*100) of the possible inputs (one bit is not applicable since at least 2-bits are required for implementing MLE circuit, one for feeding charge into source of PFET and other to drive the gate of FETs (Fig. 6). As discussed above the energy metric is greatly reduced from 2 to 3-bit (as shown in Fig.7) because less number of MLs are energized in 3-bit MLE compared to 2-bit MLE. Thus to attain the best performance in terms of overall energy metric and speed, the number of bits in MLE circuit is optimized to 3 bits. For large word sizes, greater than 32 bits, irrespective of the word length, the optimized bits for MLE circuit design remains 3. Since MLE circuit decides the number of MLs to be energized in terms of percentage as discussed above. This has been verified by us.

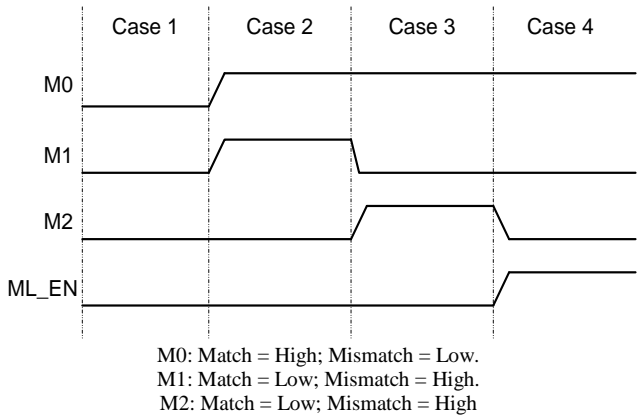

Fig. 8. Waveforms illustrating the MLE operation

*Case 1:*

If the first search bit mismatches with the stored bit in first latch, then M0 will be low else if the first search bit matches with the stored bit in first latch (S0⊕D0), then M0 will be high giving supply to the whole MLE circuit as shown in Fig. 8. In case of first bit matching, the ML_EN (Match-Line ENable) will be decided by following case 2, case 3, and case 4.

The second and third bits are matched using XOR CAM cell.

*Case 2:*

If the second bit is mismatched ($\overline{S1 \oplus D1}$), then M1 will be high. It will turn off P1 to isolate ML_EN from M0 and it will be pulled down to low through N1 as shown in Fig. 8.

*Case 3:*

If the third bit is mismatched ($\overline{S2 \oplus D2}$), the M2 will be high. It will turn off p2 to isolate ML_EN from M0 and it will be pulled down to low through N2 as shown in Fig. 8.

*Case 4:*

When both second and third bits are matching, M1 and M2 will be held low. It will turn on P1 and P2 to pull up ML_EN to high by shutting N1 and N2 as shown in Fig. 8.

## V. SELECTIVE MATCH-LINE ENERGIZER CONTENT ADDRESSABLE MEMORY (SMLE-CAM)

Core intension behind the architecture of SMLE-CAM is to energize the MLs only when the first three search bit matches before the precharge (Pre) signal goes high. SMLE-CAM architecture which is shown in Fig. 9, has Match-Line Energizer (MLE) arranged before the pass transistor ($P_0,…,P_{N-1}$) followed by NOR CAM cells. A total 253 NOR CAM cells are used per search word. In this architecture there is no critical discharge path and the MLE circuit is very fast and therefore unlike [5], implies no possible race condition problem status.

*Precharge Phase*

As shown in the timing analysis diagram of Fig. 10, the Pre control signal is held low during Precharge phase. The search bits are fed into CAM during this phase. At this point the MLE will evaluate the first three bits using the proposed XNOR and XOR CAM cells as discussed in section II. When the first three bits are successfully matched, the ML_EN will go high as discussed in section III. MLs will be precharged to high through $P_0$, $P_1$…, $P_{N-1}$ transistors whose ML_EN is held high by MLE circuit. Thus our design is very power efficient because 87.5% of MLs will not be energized in case of at least one mismatch in the first three bits considering uniform data distribution.

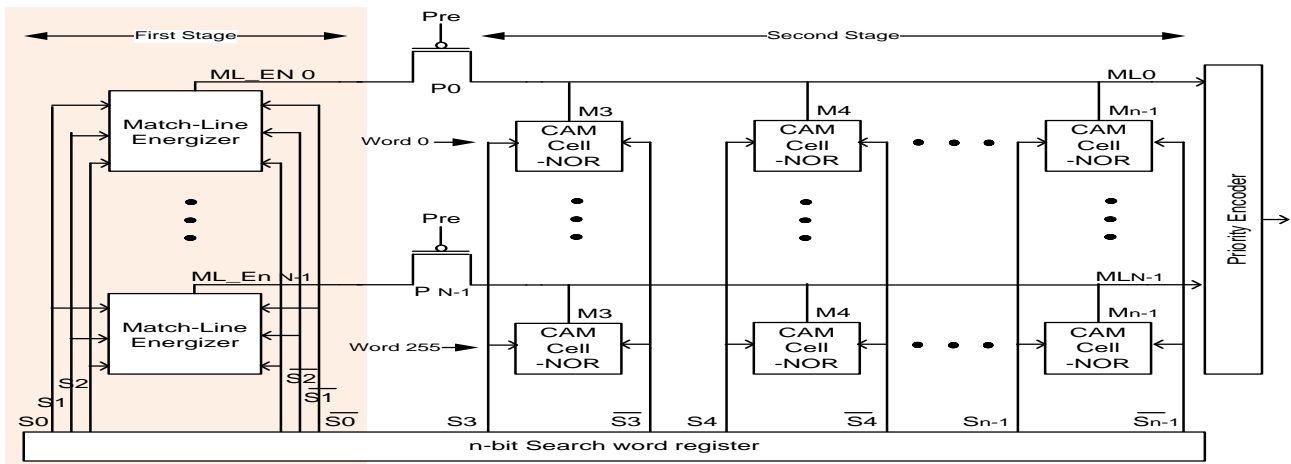

Fig. 9. Architecture of Selective Match-Line Energizer Content Addressable Memory (SMLE-CAM)

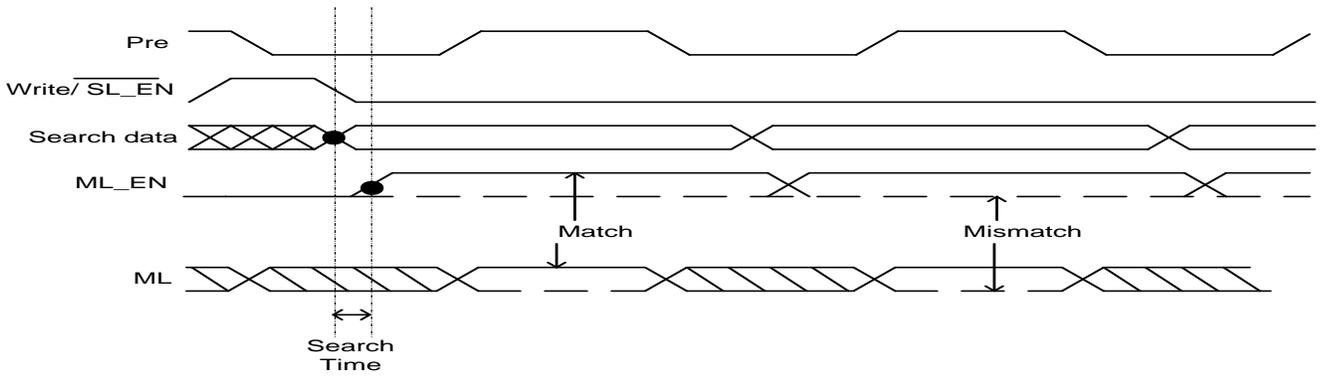
Fig. 10. Timing analysis of SMLE-CAM

*Match-line Evaluation Phase*

This phase starts when the control signal Pre starts rising i.e., high.

Case 1: Suppose the ML-EN is low as shown in Fig. 8, the ML will not be energized thus limiting power which is employed for unnecessary search in the N-3 bits (Where N is the number of search bits). The mismatch is shown by dashed lines of Fig. 10. Thus no charge will flow through pass transistors and the match-line will be low, denoting a mismatch.

If the ML_EN line is high (indicating match in first three bits as shown in Fig. 8), the particular ML will be high. Now the ML will be either held high or pulled low by the chain of NOR CAM cells as follows:

Case 2: In this case, if any of the bits is mismatched (between N-3 to N bits), then the NOR CAM cell will pull the ML low as discussed in Section II, denoting a mismatch.

Case 3: If all the bits are matched, then there will not be any pull down path and ML will remain high as in precharge phase indicating match of the word.

In our architecture, the transient between ML and raising edge of Pre is zero, thus denoting nearly zero matching delay. Thus, total search time is just the search delay as shown in Fig. 10. CAMs residing in MLE circuit use same line for placing both write and search bit. Whereas NOR CAM cells use different lines for placing write and search bit. It is to be observed from Fig. 10, when Write/$\overline{SL\_EN}$ control signal is high, write operation begins. Write bits will be placed in S0, S1 & S2 lines to write data into the CAMs placed within the MLE. Same Write/$\overline{SL\_EN}$ control signal can be fed into NOR CAM cells to write rest of the bits. The control signal will be fed to gate of the pass transistors (NMOS) for storing write bits [16] into the SRAM latch of CAM cells. During write operation, the ML_EN will have ambiguous signal. When write/$\overline{SL\_EN}$ control signal is asserted low, the ML_EN will act according to the search bits placed in S0, S1 & S2 lines.

## VI. IMPLEMENTATION RESULTS OF SMLE-CAM

Simulation results of SMLE-CAM cell in TSMC 90nm technology are shown in Fig.11. Percharging of ML_EN and ML occurs during negative level of pre signal. At first rising edge of Pre signal, ML is high indicating a match and at second rising edge of Pre signal, ML is low indicating a mismatch.

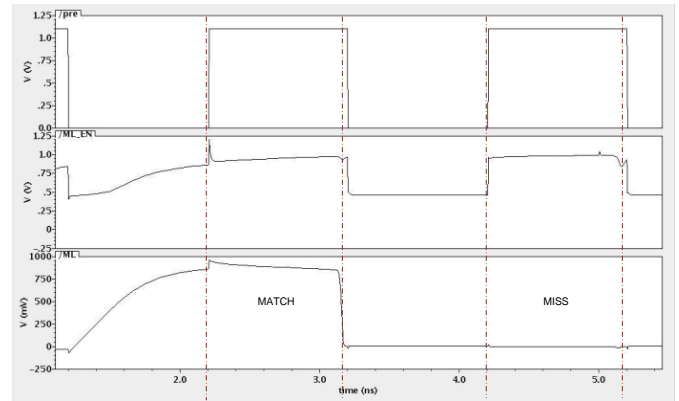
Fig. 11. Post-layout simulation result of SMLE-CAM match/miss-match

The sensitive level of match-line varies with the number of mismatch bits [14]. For more number of mismatches the ML sensitive is high and for least mismatch (a one bit) the ML sensitive is low. For this reason, those CAMs have MLSA (Match-Line Sense Amplifier) circuit, which amplifies the signal in ML to perfect match or mismatch. Unlike this in SMLE-CAM, irrespective of number of mismatching bits, ML will be held completely low as shown in Fig. 11. Thus in our architecture there is no MLSA, thereby reducing area. It is also noted that there is zero transient delay between positive edge of Pre signal and the ML, thus denoting zero ML delay.

The layout of proposed CAM is shown in Fig.12 (White borders are used to differentiate the layout areas with various functionalities).

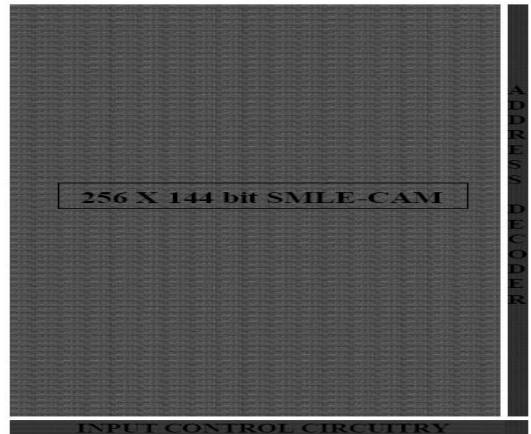
Fig. 12. Layout of Proposed SMLE-CAM

TABLE I. COMPARISON: PERFORMANCE PARAMETER OF VARIOUS CAMS

| CAM | [5] | [6] | [7] | [8] | [9] | [10] | [11] | [12] | [13] | [14] | This Work |
|---|---|---|---|---|---|---|---|---|---|---|---|
| Technology (nm) | 180 | 350 | 180 | 350 | 350 | 180 | 180 | 130 | 65 | 65 | 90 |
| Supply Voltage (V) | 1.362 | 3.3 | 1.8 | 1.5 | 3.3 | 1.8 | 1.8 | 1.0 | 1.0 | 0.5 | 1.1 |
| Word Length (bits) | 32 | 54 | 144 | 30 | 32 | 144 | 32 | 32 | 144 | 128 | 144 |
| Search Time (ns) | 0.609 | 7.3 | 4 | 4.5 | 3.9 | 7 | 2.1 | 0.9 | 0.38 | 0.75 | 0.36 |
| Energy Metric (fJ/bit/search) | 6.762 | 131 | 2.82 | 86 | 93 | 2.89 | 2.33 | 1.872 | 0.165 | 0.76 | 0.115 |

## VII. COMPARISON

Table I compares the performance of recently reported works on CAMs with the proposed. It is observed that SMLE-CAM is superior in terms of both delay and energy metric. SMLE-CAM doesn't use MLSA as discussed in section V, thus reducing area over-head along with power. The design is checked for PVT variations and data stability in the CAM cells during search operation.

The area of the CAM array implemented in 90nm node is 301415μm$^2$ for 256×144 SMLE-CAM; which is reduced by 30% compared to [13] which is implemented in 65nm node.

Hybrid CAM [5] and BMLS CAM [13] have delay of 0.609ns and 0.380ns respectively. Comparatively SMLE-CAM has 0.361ns of delay which is 5% less than BMLS CAM, since SMLE-CAM has parallel pull down path.

The BMLS CAM [13] has low energy metric of 0.165fJ/bit/search. At the worst case, our design consumes very low power, with energy metric of just 0.115 fJ/bit/search, which is 30.30% less than the BMLS CAM's energy metric. Implementation of our work in 65nm technology as done in BMLS CAM will yield even better metrics.

## VIII. CONCLUSION

SMLE-CAM architecture presents a most energy efficient and high speed CAM, by introducing Match-Line Energizer algorithm. The power efficiency of MLE circuit is coupled with the high speed search performance of NOR CAM cell. Simulation result for 144-bits search word length shows, that the energy metric of CAM is 0.115fJ/bit/search in TSMC 90nm technology. The design has 100% yield and excellent robustness against PVT variations. For future work this CAM could show splendid performance in further nanoscale CMOS technology. Also it could be interesting to enhance the robustness by adding parity bit as done in [14].